\documentclass [aps,prl,twocolumn,showpacs] {revtex4}
\usepackage[final]{graphics}
\usepackage{amssymb}
\usepackage{amsfonts}
\usepackage{epsfig}

\begin{document}

\title{Ising model on directed small-world Voronoi Delaunay random lattices}

\author{ Ediones M. Sousa and F.W.S. Lima }
\affiliation{Dietrich Stauffer Computational Physics Lab, Departamento de F\'{\i}sica,
Universidade Federal do Piau\'{\i} , 64049-550, Teresina, PI, Brazil}

\begin{abstract}

We investigate the critical properties of the Ising model
in two dimensions on {\it directed} small-world lattice with quenched connectivity
disorder. The disordered system is simulated by applying the 
Monte Carlo update heat bath algorithm.
We calculate the critical temperature, as well as the critical
exponents $\gamma/\nu$, $\beta/\nu$, 
and $1/\nu$ for several values of the rewiring probability $p$.
We find that this disorder 
system does not  belong to the same universality
class as the regular two-dimensional ferromagnetic model. The 
Ising model on {\it directed} small-world lattices presents in fact a
second-order phase transition with new critical exponents which
do not dependent of $p$, but are identical to the exponents
of the Ising model and the spin-$1$ Blume-Capel model on {\it directed} small-world network. 

\end{abstract}


\maketitle

\section{Introduction}
Experimental studies of real magnetic materials show that their critical behavior can suffer the influence of either impurities or inhomogeneities \cite{bel}.
A theoretical understanding of such impurities can be realized, with
a very good approximation, in the case of quenched disorder. In this case,
the criterion due to Harris \cite{harri} is an important theoretical tool to interpret the importance of the effect of quenched random disorder on the critical
behavior of a physical system. The randomness can be classified solely by
the specific heat exponent of the pure system, $\alpha_{pure}$. This
criterion asserts that for
$\alpha_{pure}>0$ the quenched random disorder is a relevant perturbation,
leading to a different critical behavior than in the pure case
(as for the three-dimensional Ising model). In
particular, one expects \cite{fisher} in the disordered system that
$\nu\geq 2/D$, where $\nu$ is the correlation length exponent and $D$ is the dimension 
of the system.
Assuming hyper-scaling to be valid, this implies $\alpha=2-D\nu\leq 0$. On the other hand,
for $\alpha_{pure}<0$ disorder is irrelevant (as, for instance, in the three-dimensional
Heisenberg model) and, in the marginal case
$\alpha_{pure}=0$ (like the $D=2$ Ising model), no prediction can be made.
For the case of (non-critical) first-order phase transitions it is known that the influence of
quenched random disorder can lead to a softening of the transition \cite{Imry}.
Recently,  the predicted softening effect at first-order phase transitions
has been confirmed for D=3 q-state Potts models with $q\geq3 $ using 
Monte Carlo \cite{balestro,chatelain1,chatelain2} and high temperature series
expansion \cite{M} techniques.
The overall picture is even better in two dimensions ($D=2$) where several models
with $\alpha_{pure}>0$ \cite{D,Ma,Gj,SW} and the marginal ($\alpha_{pure}=0$) 
\cite{25,28,aarao,puli,puli2} have been investigated.

In this paper we study a type of different quenched disorder, namely the effect of
{\it directed} bound case with rewiring probability $p$ \cite{sanches}.
Specifically, we consider $D=2$ small-world Voronoi Delaunay random lattices (SWVD) type, 
and performed an extensive computer simulation study of Ising model.
We concentrated on the close vicinity of the transition point
and applied finite-size scaling (FSS) techniques to extract the exponents and the
``renormalized charges" $U_{4}^{*}$. Monte Carlo simulations of the disorder system was realized using the spin-flip heat bath algorithm to update 
the spins.
Previous studies of connectivity disorder on $D=2$ lattices have been realized 
by Monte Carlo simulations of $q$-state Potts models on quenched random lattices of 
Voronoi Delaunay type for $q=2$ \cite{janke,janke1,FWSL}, $q=3$ \cite{FWSL1} and $q=8$ 
\cite{janke2,FWSL2}. In particular, it has been shown that for  $q=2$ 
\cite{janke,janke1,FWSL} and $q=3$  \cite{FWSL1} the critical
exponents are the same as those for the  model on a regular $D=2$ lattice.
This is indeed a surprising result since the relevant criterion of the Delaunay triangulations 
reduces to the well known Harris criterion such that disorder of this type should be relevant
for any model with positive specific heat exponent \cite{JankeWeigel}. This means that  
for $q=3$, where  $\alpha_{pure}>0$,  one would expect a different universality class. 
For the spin-1 Ising model, where  $\alpha_{pure}=0$, 
Fernandes et al. \cite{fernandes}  showed that the exponents do no change in the {\it undirected} SWVD lattice, but for {\it directed} SWVD random lattice the situation is quite different. There is a second-order phase transition for $p<p_{c}$ and a first-order phase transition for $p>p_{c}$, where $p_{c}\approx 0.35$ is the rewiring probability where the system change the order phase transition. In addition,
the calculated critical exponents for
$p<p_{c}$ do
not belong to the same universality
class as the regular two-dimensional ferromagnetic model. Therefore both {\it undirected} and {\it directed} cases agree with Harris criterion for $\alpha_{pure}=0$. 

In the present spin-$1/2$ Ising model on {\it directed} SWVD lattice we show that the critical behavior is quite similar to that observed by Fernandes et al. in the spin-$1$ case\cite{fernandes}. However, now one has only a second-order phase transition for all $p$ values studied. The critical exponents do
not belong to the same universality
class as the regular two-dimensional ferromagnetic model, but they agree with the critical exponents of Blume-Capel model for 
$p<p_{c}$ \cite{fernandes}.
In the next section we present the 
model and the simulation background. The results and
conclusions are discussed in the last section.
\bigskip
 
\section{Model and  Simulation}

We consider the ferromagnetic spin-$1/2$ Ising model, on {\it directed} SWVD random lattice by a set of spins variables ${S_i}$ taking the values $\pm 1$ situated on every site $i$ of a {\it directed} 
SWVD random lattice with $N=L\times L$ sites, were $L$ is the side of square cluster. In this random lattice, similar to 
S\'anchez et al. \cite{sanches}, we start from a two-dimensional SWVD random
lattice consisting of sites linked to their $c$ ( where $3<c<20$ and different for each site of network) nearest neighbors by both outgoing and
incoming links. Then, with probability $p$, we reconnect nearest-neighbor outgoing
links to a different site chosen at random. After repeating this process for every
link, we are left with a network with a density $p$ of SWVD {\it directed} links. Therefore,
with this procedure every site will have $c$ outgoing links and
varying (random) number of incoming links. 

The evolution in time of these systems is given by a single spin-flip
like dynamics with a probability $P_{i}$ given by
\begin{equation}
P_{i}= 1/[1+\exp(2E_{i}/k_BT)], 
\end{equation}
where $T$ is the temperature, $k_B$ is the Boltzmann constant, and
$E_i$ is the energy of the configuration obtained from the
Hamiltonian
\begin{equation}
H=-J\sum_{<i,j>}S_{i}S_{j},
\end{equation}
where the sum runs over all neighbor pairs of sites  
(including the nearest-neighbor and the long ranged ones determined by the probability $p$)
and the spin-$1/2$ variables $S_{i}$ assume values 
$\pm 1$. In the above equation $J$ is the exchange coupling. The spin-$1/2$ case is 
well known in the literature \cite{a3, onsager}. 

The simulations have been performed on  
different SWVD random lattice sizes  comprising a number $N=5000$, $10000$, $20000$, $40000$, $60000$ and $80000$ 
of sites. For simplicity,  the length of the system is defined here in terms
of the size of a regular lattice $L=N^{1/2}$.
For each system size quenched averages over the connectivity disorder are 
approximated by averaging over $R=20$ independent realizations. For each
simulation we have started with a uniform configuration of spins. 
We ran $4\times10^{5}$ Monte Carlo steps (MCS) per spin with $2\times10^{5}$ configurations 
discarded for thermalization using the ``perfect" random-number generator \cite{nu}. 
We do not see any significant change by increasing
the number of R and MCS. So, for the sake of saving computer time, the
present values seem to give reasonable results for our simulation.

In both cases we have employed the heat bath algorithm and
for every MCS, the energy per spin,
$e=E/N$, and the magnetization per spin, $m=\sum_{i}S_{i}/N$, were
measured.
From the energy measurements we can compute the average energy, specific heat and
the fourth-order Binder cumulant of the energy, given respectively by
%
\begin{equation}
 u(T)=[<E>]_{av}/N,
\end{equation}
\begin{equation}
 C(T)=K^{2}N[<e^{2}>-<e>^{2}]_{av},
\end{equation}
\begin{equation}
 B(T)=1-[\frac{<e^{4}>}{3<e^{2}>^{2}}]_{av}.
\end{equation}
In the above equations
$<...>$ stands for  thermodynamic averages and $[...]_{av}$
for  averages over different realizations.
Similarly, we can derive from the magnetization measurements
the average magnetization, the susceptibility, and the fourth-order magnetic cumulant,
\begin{equation}
 m(T)=[<|m|>]_{av},
\end{equation}
\begin{equation}
 \chi(T)=KN[<m^{2}>-<|m|>^{2}]_{av},
\end{equation}
\begin{equation}
 U_{4}(T)=1-[\frac{<m^{4}>}{3<|m|>^{2}}]_{av}.
\end{equation}

In order to calculate the exponents of this model, we apply finite-size scaling
(FSS) theory. We
then expect, for large system sizes, an asymptotic FSS behavior of the form
\begin{equation}
 C=C_{reg}+L^{\alpha/\nu}f_{C}(x)[1+...],
\end{equation}
\begin{equation}
 [<|m|>]_{av}=L^{-\beta/\nu}f_{m}(x)[1+...],
\end{equation}
\begin{equation}
 \chi=L^{\gamma/\nu}f_{\chi}(x)[1+...],
\end{equation}
where $C_{reg}$ is a regular background term, 
 $\nu$, $\alpha$, $\beta$, and $\gamma$ are the usual critical
exponents, and $f_{i}(x)$ are FSS functions with
\begin{equation}
 x=(T-T_{c})L^{1/\nu}
\end{equation}
being the scaling variable. The dots in the brackets $[1+...]$ indicate
corrections-to-scaling terms. We calculated the error bars from the fluctuations 
among the different realizations. Note that these errors contain both, the average
thermodynamic error for a given realization and the theoretical
variance for infinitely accurate thermodynamic averages which are
caused by the variation of the quenched, random geometry of the lattices.

\section{Results and conclusion}
By applying the standard heat bath algorithm to each of the $R$ energy data
we determine the temperature dependence of $C_{i}(T)$, $\chi_{i}(T)$,..., $i=1$,...,$R$.
Once the temperature dependence is known for each realization, we can easily compute
the disorder average, e.g., $C(T)=\sum^{T}_{i=1}C_{i}(T)/R$, and then determine the maxima 
of the averaged quantities, e.g., $C_{max}(T_{max})=max_{T}C(T)$. The variable $R$ ($=20$)
represents the number of replicas in our simulations.

\begin{figure}[ht]
\includegraphics[clip,angle=0,width=8.0cm]{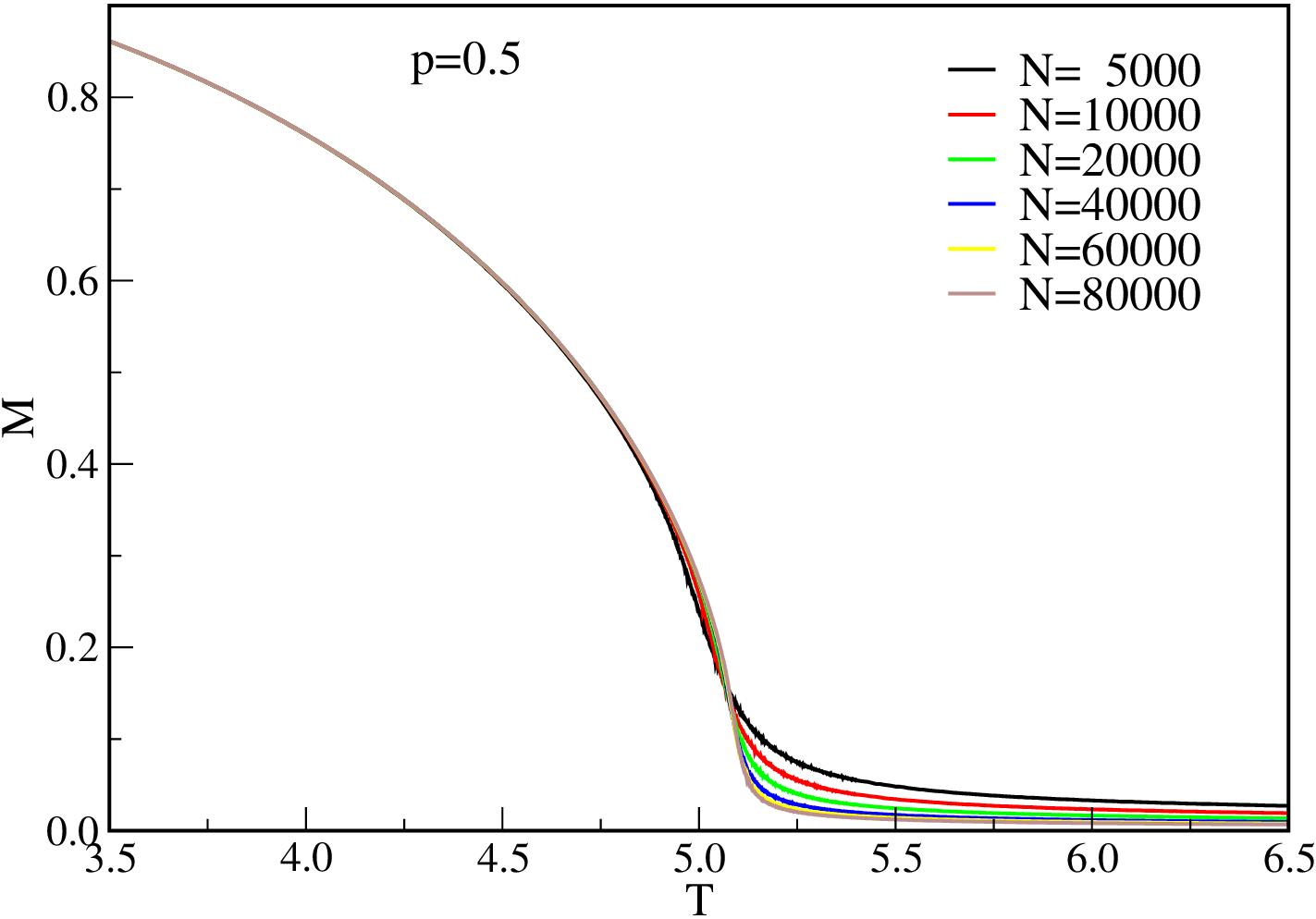}
\caption{\label{cum0} (color online) Magnetization as a function of $T$ for various
lattice sizes with $N=5000$, $N=10000$, $N=20000$, $N=40000$, $N=60000$, and
$N=80000$ and rewiring probability $p=0.5$.}
\end{figure}

\begin{figure}[ht]
\includegraphics[clip,angle=0,width=8.0cm]{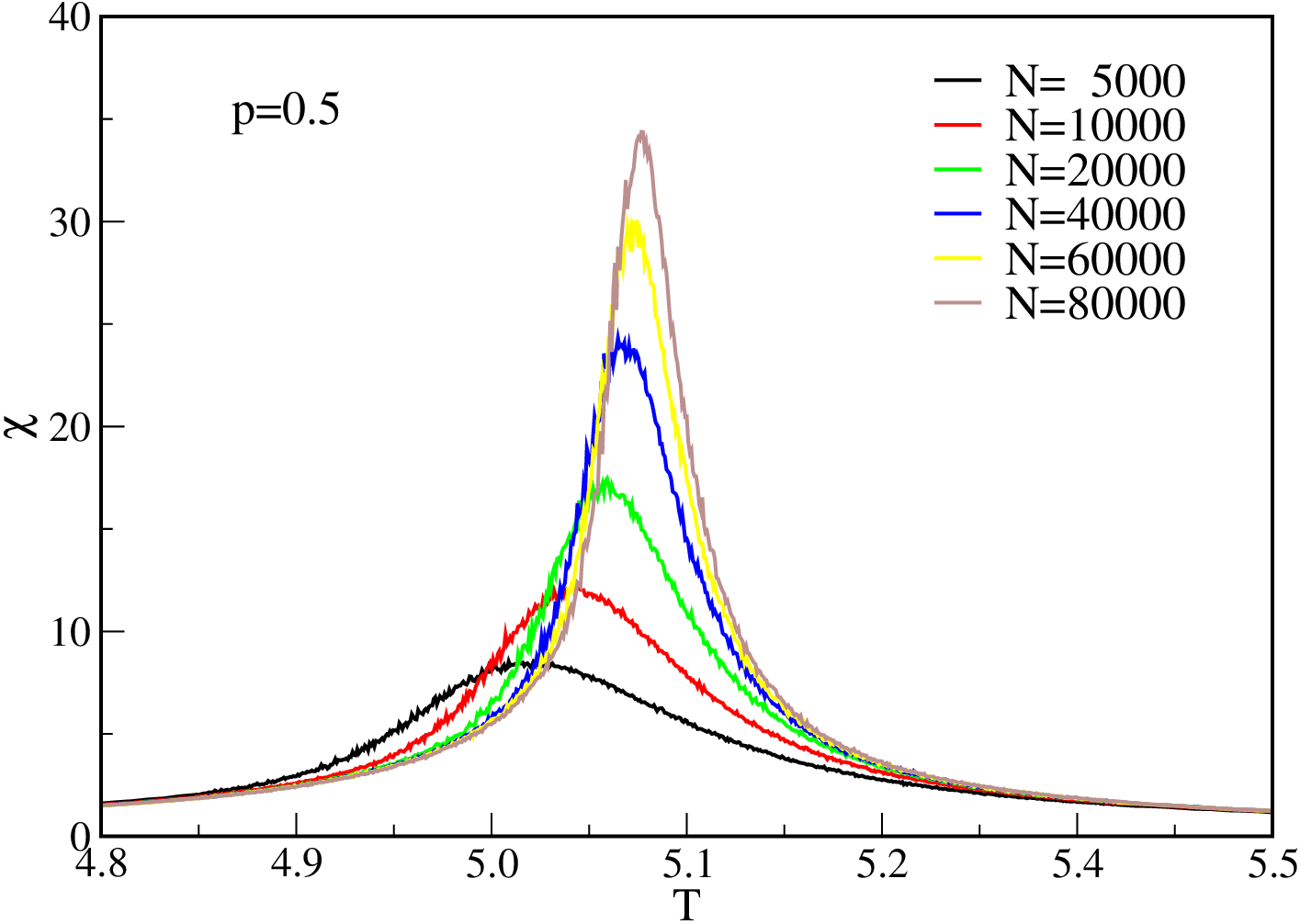}
\caption{\label{cum1} (color online)
The same as Figure \ref{cum0} for the susceptibility versus temperature $T$.}
\end{figure}

\begin{figure}[ht]
\includegraphics[clip,angle=0,width=8.0cm]{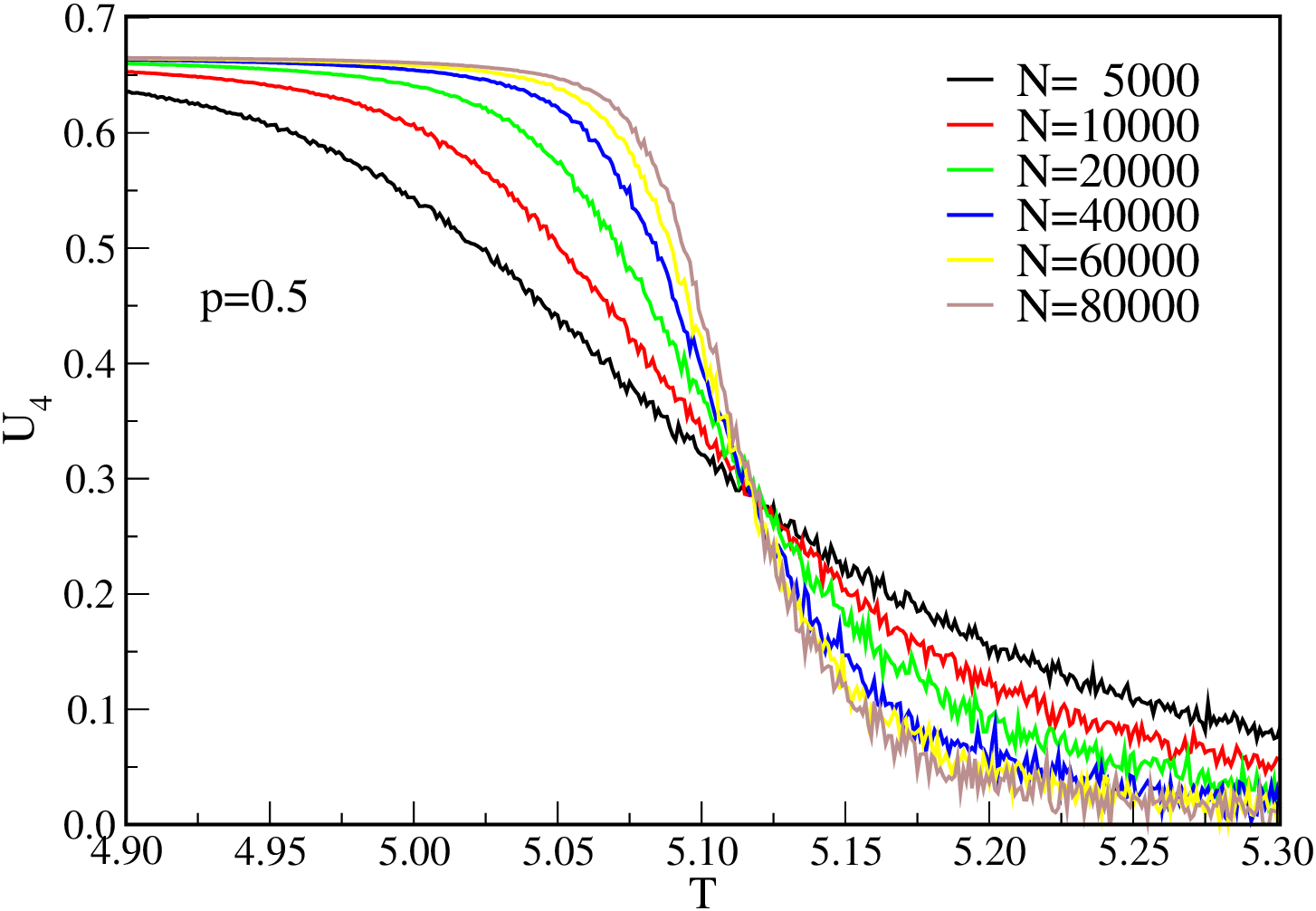}
\caption{\label{cum2} (color online) The same as Figure \ref{cum0} for the
fourth-order Binder cumulant as a function of $T$.}
\end{figure}

In Figure \ref{cum0} we show the behavior of the magnetization versus temperature
for  several different lattice sizes and rewiring probability $p=0.5$. Figure \ref{cum1} displays the behavior of the suscepbility versus temperature for the same parameters used in Figure \ref{cum0}. From here on we set $J$ and $k_B$ to unity. One can see a typical behavior of a 
second-order phase transition. In  order to estimate the critical temperature we calculate the fourth-order 
Binder cumulant given by  eq. (8). It is well known that these
quantities are independent of the system size and should intercept at the critical 
temperature \cite{binder}. In Figure \ref{cum2} the fourth-order
Binder cumulant is shown as a function of the $T$ for several lattice sizes
for the rewiring probability $p=0.5$. Taking the 
largest lattices we have  $T_{c}= 5.118(4)$. To estimate $U^{*}_{4}$ we note that it
varies little at  $T_{c}$, 
so we have $U^{*}_{4}= 0.283(4)$. One can see that $U^{*}_{4}$ is 
different from the universal value $U^{*}_{4}\sim 0.61$  for the Ising model on the regular $D=2$ lattice, and also for the Ising model on Voronoi-Delaunay random lattice in two-dimensions \cite{janke,janke1,FWSL}. By following this same procedure one can
get the corresponding results for other values of $p$.
%
%
%
%
%
%
%

\begin{figure}[ht]
\includegraphics[clip,angle=0,width=8.0cm]{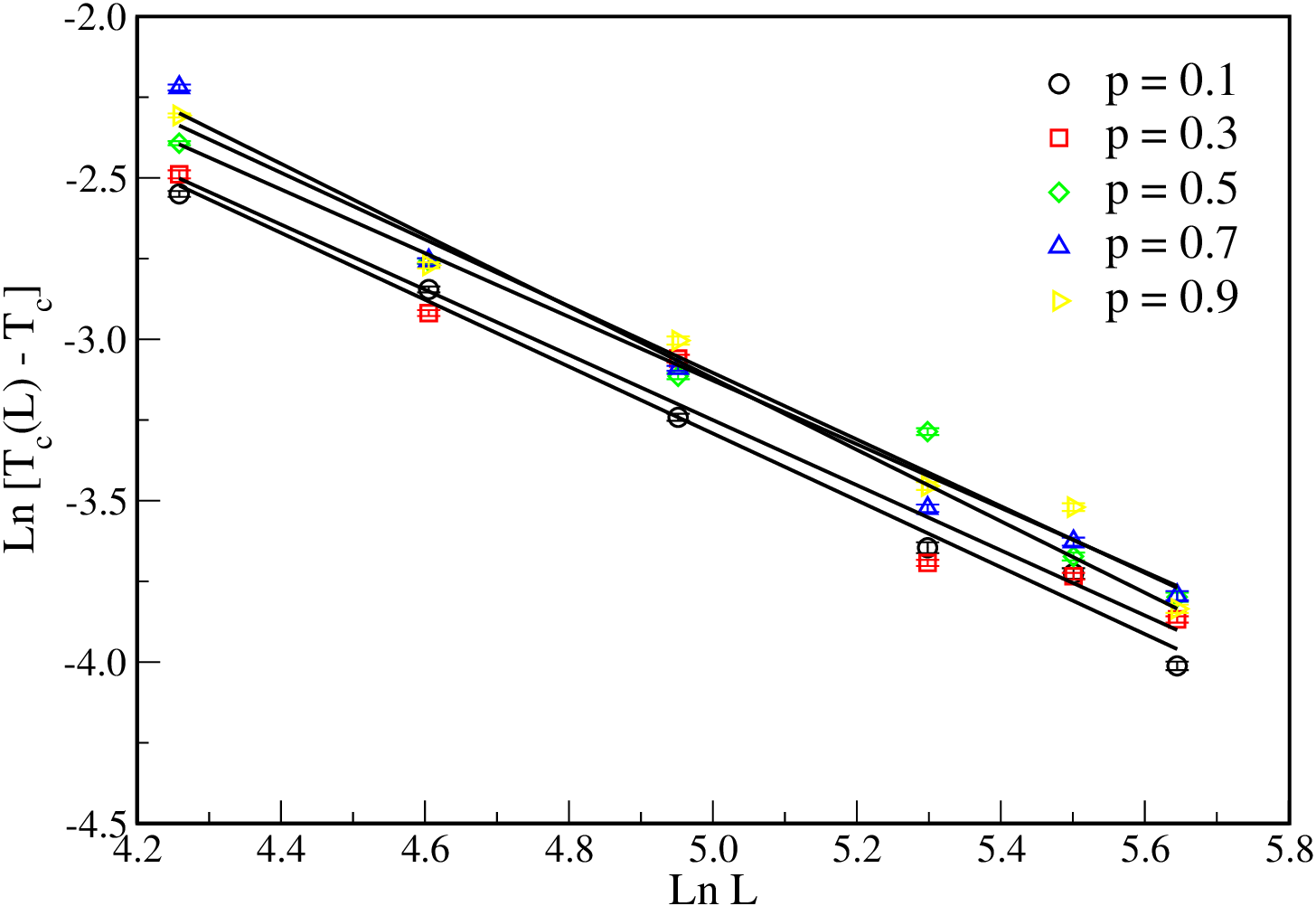}
\caption{\label{expnu}(color online)
 $ \ln\left[T_{c}(L)-T_{c}\right]$ as a function of $L$ for several values of $p$. The solid lines are the best linear fits. 
}
\end{figure}

The correlation length exponent can be estimated from
$ T_{c}(L)=T_{c}+bL^{-1/\nu}$, where $T_c(L)$ is the pseudo-critical temperature
for the lattice size $L$, $T_c$ is the critical temperature in the thermodynamic limit, and $b$ is a non-universal constant. In Figure \ref{expnu} it is shown a plot of $\ln\left[T_c(L)-T_c\right]$ as a function of $\ln L$ for several values of $p$.
One can clearly see that the exponent is, within the errors, independent of $p$, in agreement with universality ideas. The actual values of $1/\nu$ are displayed in Table I.

%
%
%

\begin{figure}[ht]
\includegraphics[clip,angle=0,width=8.0cm]{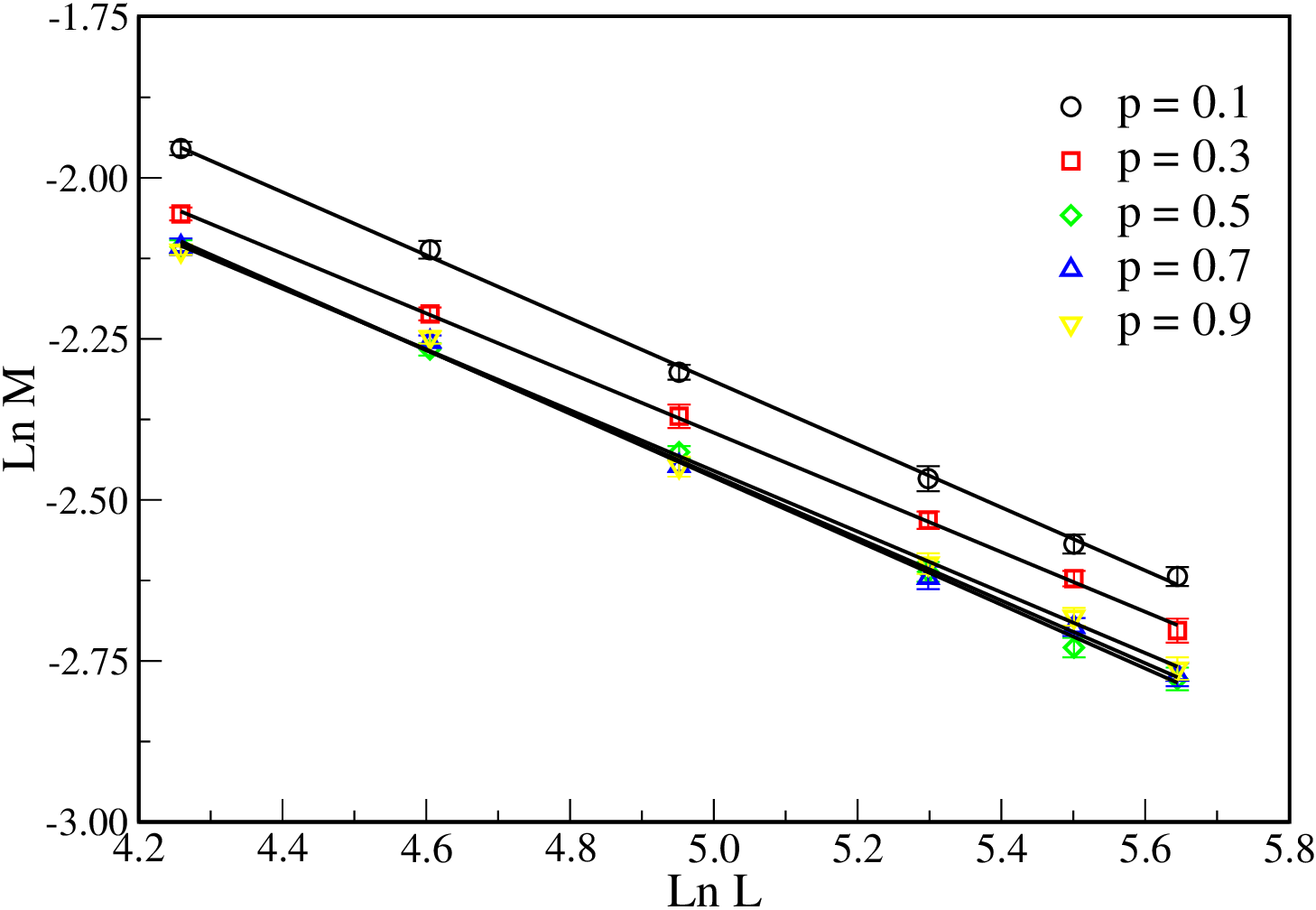}
\caption{\label{mag} (color online) 
Plot of the logarithm of the modulus of the magnetization at the
inflection point as a function of the logarithm of  $L$. The solid lines are 
the best linear fit.
}
\end{figure}

\begin{figure}[ht]
\includegraphics[clip,angle=0,width=7.9cm]{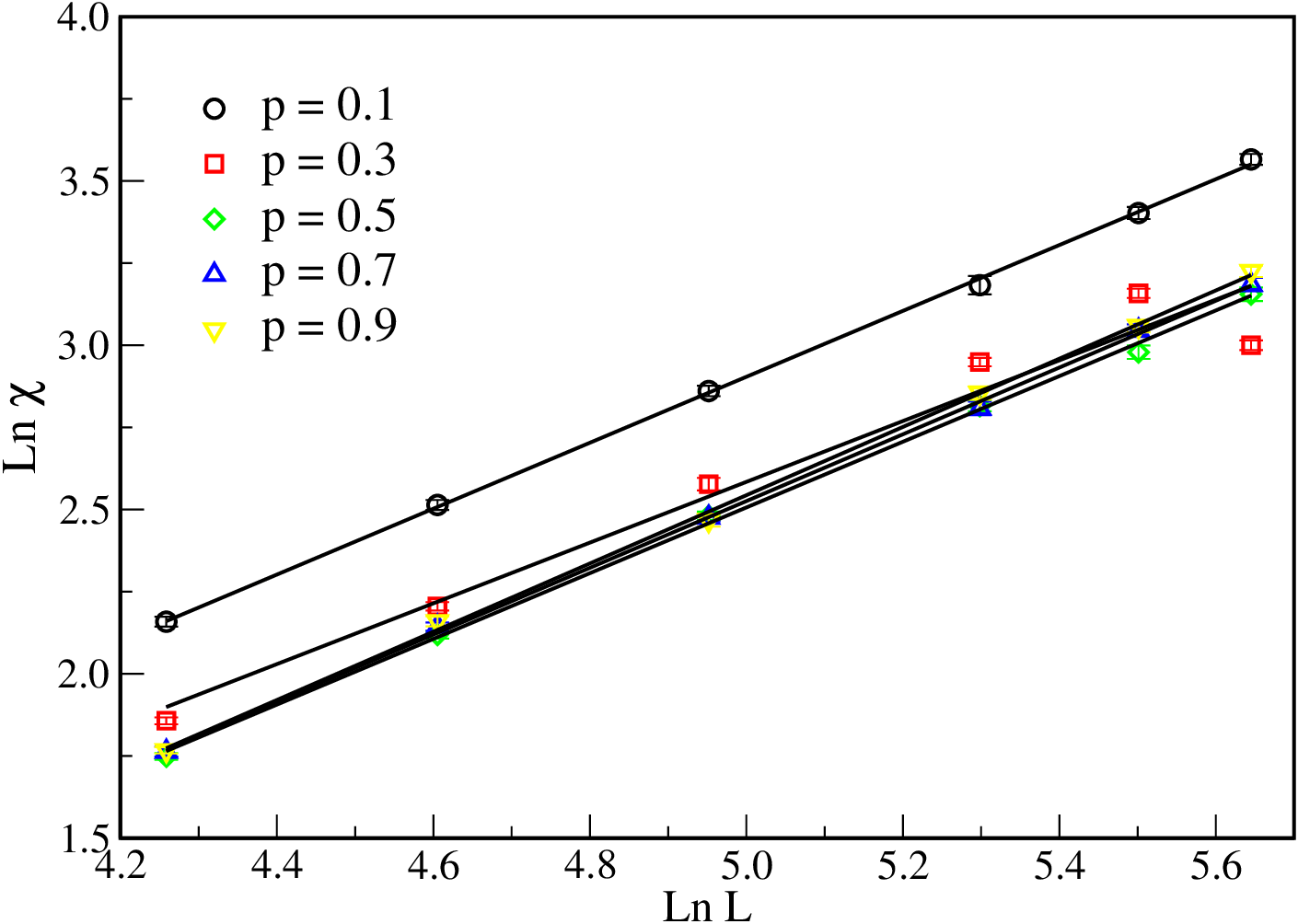}
\caption{\label{sus} (color online) 
 Log-log plot of the susceptibility $\chi$  at $T_{c}$ as a function of
the logarithm of  $L$. The solid lines are the best linear fit.
}
\end{figure}

\begin{figure}[ht]
\includegraphics[clip,angle=0,width=8.0cm]{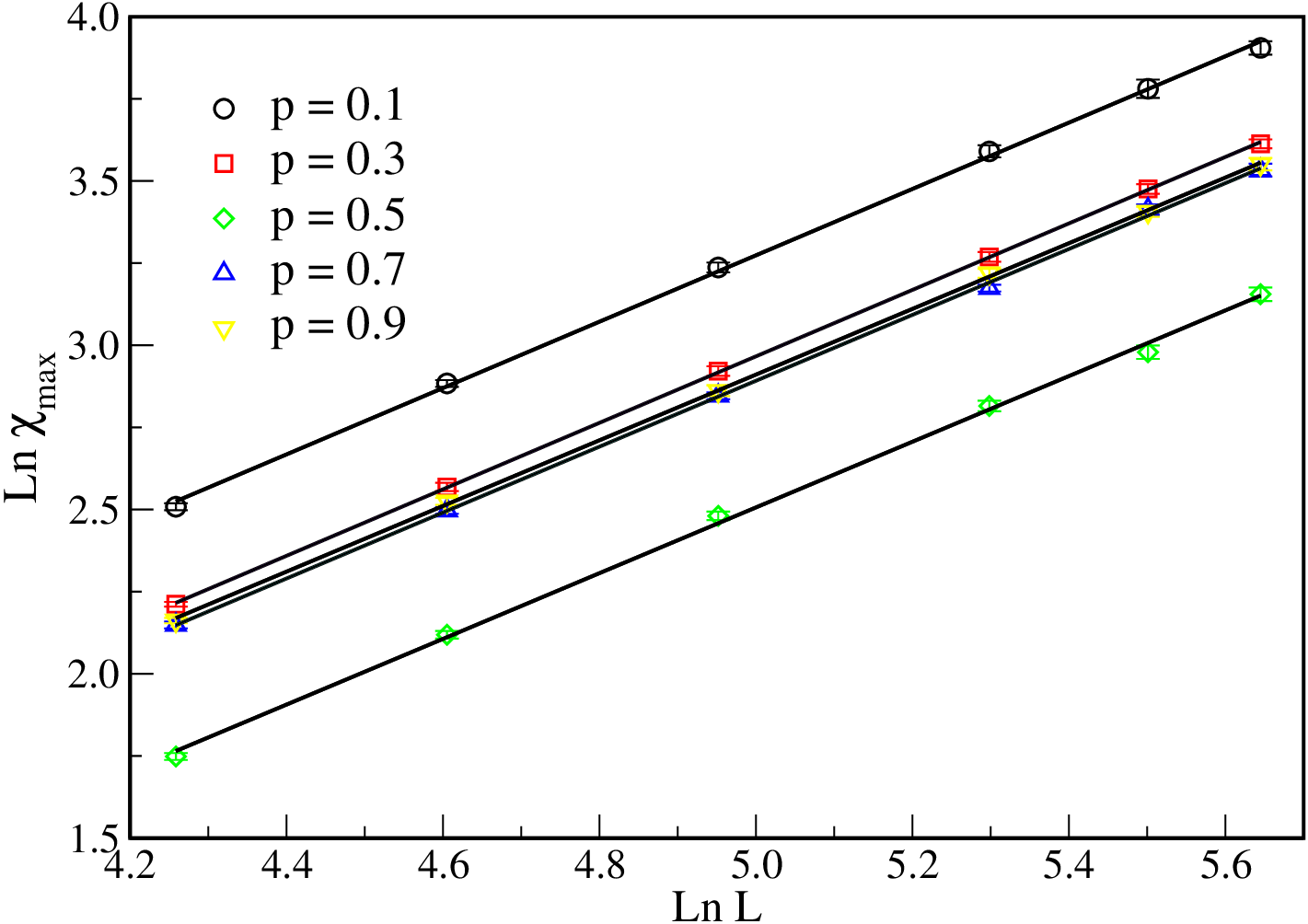}
\caption{\label{mag1} (color online) 
 Log-log plot of the susceptibility maxima $\chi_{max}$ as a function of
the logarithm of  $L$. The solid lines are the best linear fit.}
\end{figure}

In order to go further in the present analysis we have also computed the modulus of the magnetization
at the inflection point and the magnetic susceptibility at $T_c$. The logarithm of
these quantities as a function of the logarithm of  $L$ are presented in Figures  
\ref{mag} and \ref{sus},
respectively. A linear fit of these data gives  $\beta/\nu$ from the magnetization
and  $\gamma/\nu$ from the susceptibility. In addition, we plotted in Figure \ref{mag1} the logarithm of the maximum value of the susceptibility $\chi_{max}$ as a function of $\ln L$ for several values of $p$. One can also see that the exponents $\beta/\nu$ and $\gamma/\nu$ are also independent of $p$, as expected.
They are different from
$\beta/\nu=0.125$ and  $\gamma/\nu=1.75$ obtained for a regular $D=2$ lattice,
but obey hiper-scaling relation (into the error bar)
\begin{equation}
 2\frac{\beta}{\nu}+ \frac{\gamma}{\nu}=D,
\end{equation}
where $D=2$. The numerical values of the ratio $\beta/\nu$ and  $\gamma/\nu$
are also shown in Table I.
%
%
%
%

\begin{figure}[ht]
\includegraphics[clip,angle=0,width=8.0cm]{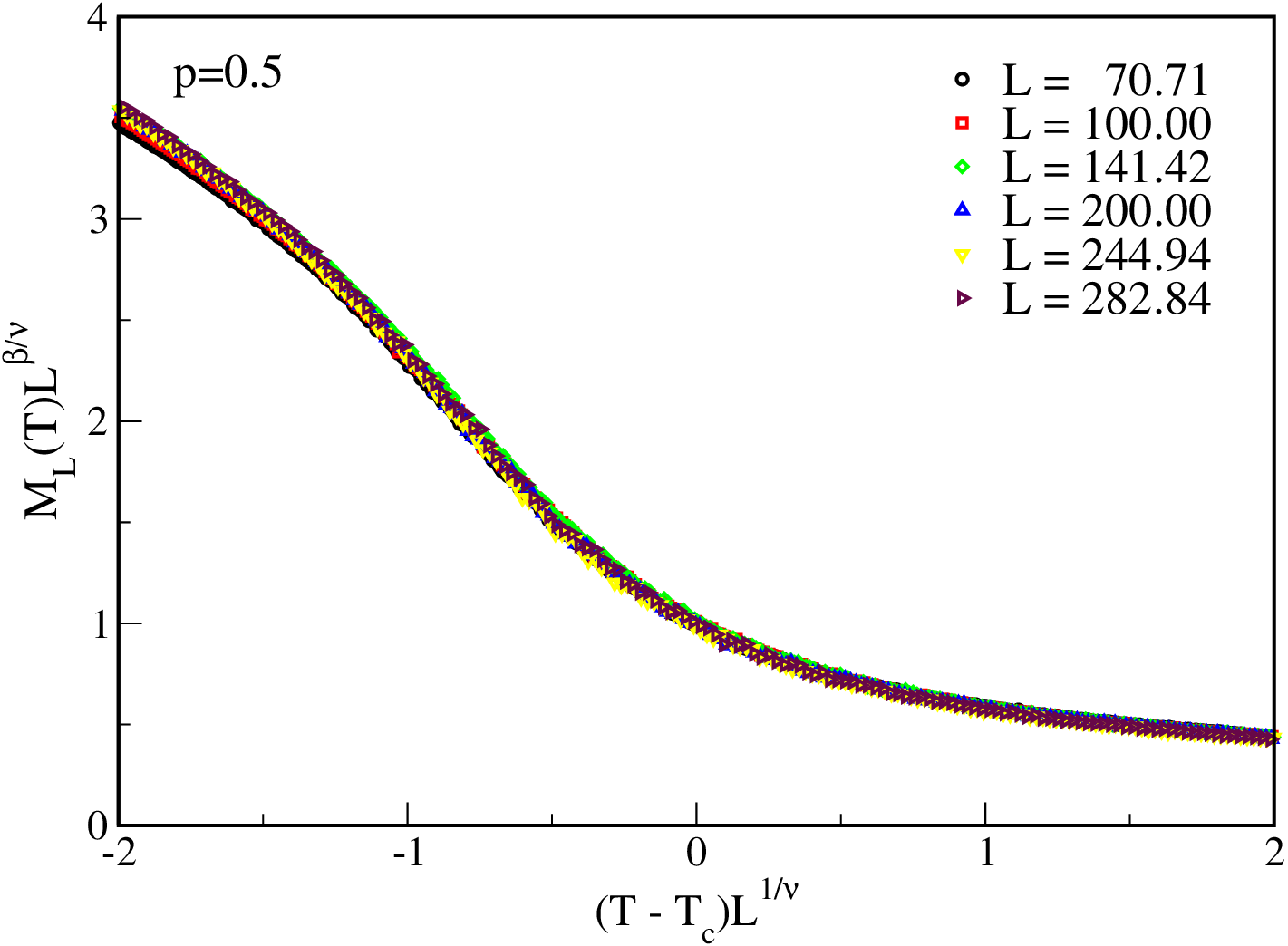}
\caption{\label{cola1} (color online) 
Data colapse of magnetisation for various values of $L$ and $p=0.5$.}
\end{figure}

\begin{figure}[ht]
\includegraphics[clip,angle=0,width=8.0cm]{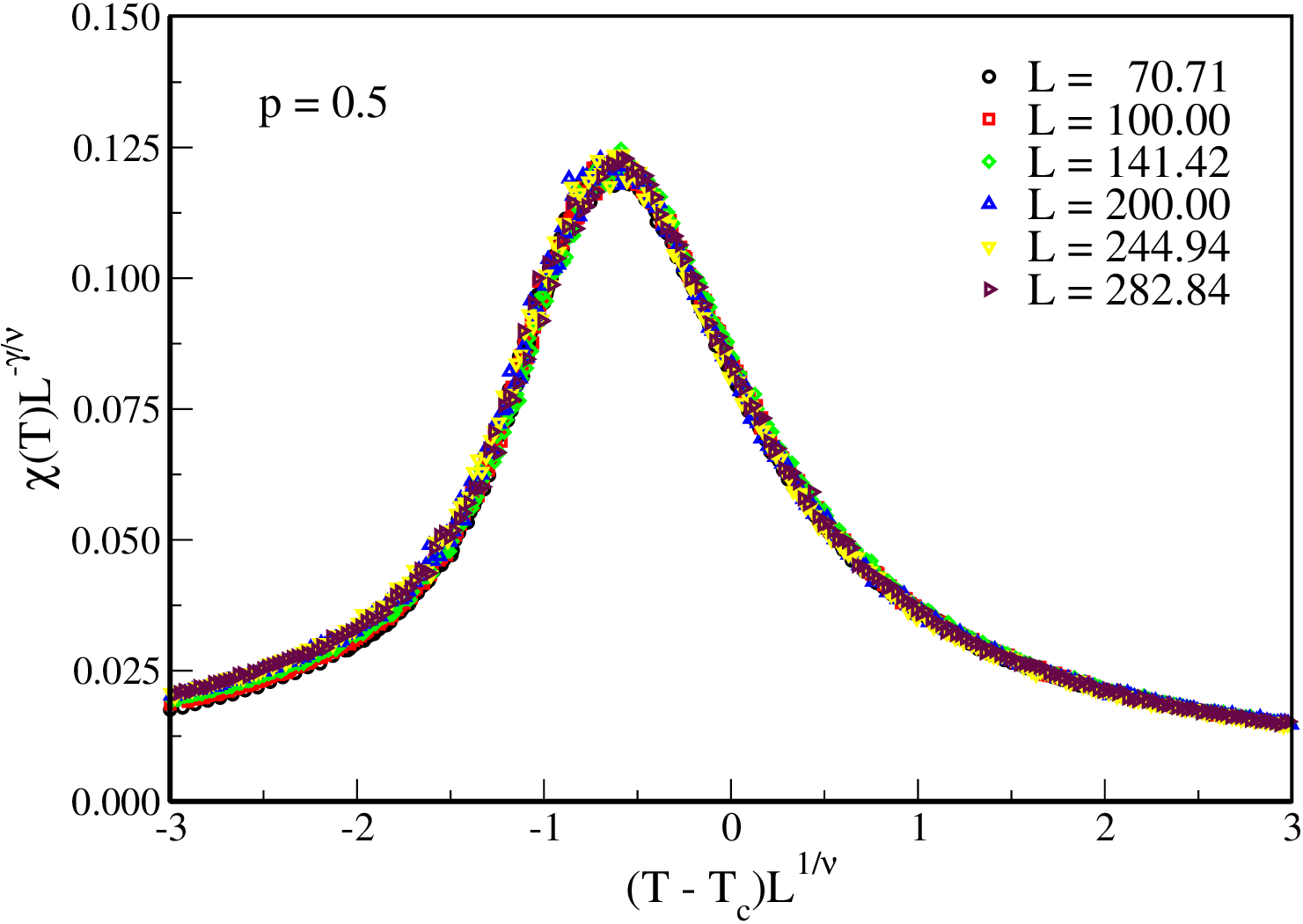}
\caption{\label{cola2}
The same as in Figure \ref{cola1} for the susceptibility.}
\end{figure}

In Figures \ref{cola1} and \ref{cola2} we display the data colapse for the magnetisation and the susceptibility
for $p=0.5$. In these cases, we see that the estimative of the critical exponents ratio $\beta/\nu$ and
$\gamma/\nu$ are in good agreement for all lattice sizes.  The same qualitative
results are obtained for other values of $p$.
%
\begin{table}[t]
\begin{center}
\caption{The critical exponents, 
for  spin-$1/2$ on {\it directed} SWVD random lattice with probability $p$. 
$\gamma/\nu^{max}$ are the results from the maximum of the magnetic susceptibility. Error bars are statistical only.} 
\begin{tabular}{|c||c|c|c|c|c|c|}
\hline
$p$   & $1/\nu$ &  $\beta/\nu$ & $\gamma/\nu$ & $\gamma/\nu^{max}$ \\
\hline
$0.1$ & $1.036(49)$ & $0.489(8)$  & $1.003(11)$ & $1.001(13)$\\
\hline 
$0.2$ & $1.098(82)$ & $0.538(68)$ & $1.016(11)$ & $1.016(5)$\\
\hline 
$0.3$ & $1.009(49)$ & $0.463 (4)$ & $0.924(98)$ & $1.012(3)$\\
\hline
$0.4$ & $0.886(8)$  & $0.491 (9)$ & $1.017(14)$ & $1.012(8)$\\
\hline
$0.5$ & $0.987(64)$ & $0.494(10)$ & $0.998(18)$ & $1.005(66)$\\
\hline
$0.6$ & $0.927(92)$ & $0.486(10)$ & $1.042(13)$ & $1.004(7)$\\
\hline
$0.7$ & $1.107(60)$ & $0.486(10)$ & $1.016(13)$ & $1.003(10)$\\
\hline
$0.8$ & $0.972(57)$ & $0.493(16)$ & $1.018(23)$ & $1.021(7)$\\
\hline
$0.9$ & $1.032(66)$ & $0.471(12)$ & $1.038(16)$ & $0.991(69)$\\
\hline
\end{tabular}
\end{center}
\end{table}

In summary, from the above results, there is a strong indication that the spin-$1/2$ Ising model 
model on a {\it directed} SWVD random lattice is in a different universality class than  the model on a regular two-dimensional lattice. The exponents here obtained are independent of $p$ and different from the Ising model on regular $D=2$ lattice,
but they are equivalent to the exponents
of the Ising model and the spin-$1$ Blume-Capel model on {\it directed} small-world network \cite{newpla}. One possible explanation for this change in universality can be ascribed to the influence of long range interactions that occur with the presence of $p$ directed bounds.  However, our results agree with the Harris-Luck
criterion for {\it directed} SWVD random lattice.

\end{document}